\documentclass[preprint,superscriptaddress]{revtex4-1}
\usepackage[utf8]{inputenc}
\usepackage{enumitem}
\usepackage{amsfonts}
\usepackage{amssymb}
\usepackage{dsfont}
\usepackage{color}
\usepackage{graphicx}
\usepackage{textcomp}
\usepackage{cancel}
\usepackage{dcolumn}
\usepackage{bm}
\usepackage{multirow}
\usepackage{float}
\usepackage{mathrsfs} 
\usepackage{braket}
\usepackage{amsmath}
\usepackage{subfig}

\begin{document}
\title {Generating scalable entanglement of ultracold bosons in superlattices through resonant shaking}

\author{Lushuai Cao}
\email{lushuai\_cao@hust.edu.cn}
\affiliation{MOE Key Laboratory of Fundamental Physical Quantities Measurement,
Hubei Key Laboratory of Gravitation and Quantum Physics,
School of Physics, Huazhong University of Science and Technology,
Wuhan 430074, People's Republic of China}
\author{Xing Deng}
\affiliation{MOE Key Laboratory of Fundamental Physical Quantities Measurement,
Hubei Key Laboratory of Gravitation and Quantum Physics,
School of Physics, Huazhong University of Science and Technology,
Wuhan 430074, People's Republic of China}
\author{Qian-Ru Zhu}
\affiliation{MOE Key Laboratory of Fundamental Physical Quantities Measurement,
Hubei Key Laboratory of Gravitation and Quantum Physics,
School of Physics, Huazhong University of Science and Technology,
Wuhan 430074, People's Republic of China}
\author{Xiao-Fan Xu}
\affiliation{Institute of Space-based Information System, China Academy of Electronics and Information Technology,
Beijing 100041, People's Republic of China}
\author{Xue-Ting Fang}
\affiliation{MOE Key Laboratory of Fundamental Physical Quantities Measurement,
Hubei Key Laboratory of Gravitation and Quantum Physics,
School of Physics, Huazhong University of Science and Technology,
Wuhan 430074, People's Republic of China}
\author{Xiang Gao}
\affiliation{MOE Key Laboratory of Fundamental Physical Quantities Measurement,
Hubei Key Laboratory of Gravitation and Quantum Physics,
School of Physics, Huazhong University of Science and Technology,
Wuhan 430074, People's Republic of China}
\author{Peter Schmelcher}
\affiliation{Zentrum für Optische Quantentechnologien, Universität Hamburg, Luruper Chaussee 149, 22761 Hamburg, Germany}
\affiliation{The Hamburg Centre for Ultrafast Imaging, Luruper Chaussee 149, 22761 Hamburg, Germany}
\author{Zhong-Kun Hu}
\email{zkhu@hust.edu.cn}
\affiliation{MOE Key Laboratory of Fundamental Physical Quantities Measurement,
Hubei Key Laboratory of Gravitation and Quantum Physics,
School of Physics, Huazhong University of Science and Technology,
Wuhan 430074, People's Republic of China}

\begin{abstract}
Based on a one-dimensional double-well superlattice with a unit filling of ultracold atoms per site,
we propose a scheme to generate scalable entangled states in the superlattice
through resonant lattice shakings. Our scheme utilizes periodic lattice modulations to
entangle two atoms in each unit cell with respect to their orbital degree of freedom,
and the complete atomic system in the superlattice becomes a cluster of bipartite entangled atom pairs.
To demonstrate this we perform $ab \ initio$ quantum dynamical simulations using the Multi-Layer
Multi-Configuration Time-Dependent Hartree Method for Bosons, which accounts for all correlations
among the atoms.
The proposed clusters of bipartite entanglements manifest as an essential
resource for various quantum applications, such as measurement based quantum computation.
The lattice shaking scheme to generate this cluster possesses advantages such as
a high scalability, fast processing speed, rich controllability on the target entangled states,
and accessibility with current experimental techniques.
\end{abstract}
\maketitle

\textit{Introduction.\textemdash} Ultracold atoms in optical lattices, that benefit from their almost perfect
decoupling from the environment and their excellent tunability \cite{Bloch2005,Bloch2008c},
have become a promising platform for realizing quantum
entanglement with related applications, such as quantum metrology \cite{entanglement_measurement},
and quantum computation \cite{original_QC,original_QC2}.
There have been various protocols for quantum computing with ultracold atoms, among which the
measurement-based quantum computation (MBQC) \cite{MBQC} is particularly suitable for lattice atoms.
MBQC is incorporated with two elementary steps, including the
preparation of scalable multiparticle entangled states \cite{cluster_state,cluster} and
operations of local quantum gates on the entangled states.
Various generation schemes have been proposed or even experimentally realized for multiparticle entangled states,
such as those based on controlled collisions \cite{original_QC,ccollision_original,controlled_collision}
and super-exchange interactions \cite{superexchange_original,ccollision,superexchange,Jaksch,tweezer,dw_entanglement},
among which the entanglement is mainly encoded into the internal degrees of freedom (DoF) of the atoms.
There is now also a growing interest to use the orbital DoF \cite{porbital_review} of lattice atoms for quantum computations,
which possesses advantages such as insensitivity to external magnetic fluctuations and
high controllability by lattice modulations.
Various quantum gate operations have been proposed for the orbital DoF
\cite{porbital_2002,porbital_2002_2,porbital_2008,interaction_blockade,porbital_interm}, and it
demands an efficient scheme to generate scalable multiparticle entangled states with respect to the orbital
DoF, to accomplish a complete MBQC.

The orbital DoF has a relatively short coherence time, mainly due to the spontaneous decay of higher orbitals,
and schemes involving second-order hoppings become unsuitable for generating entanglement with respect to
the orbital DoF. Recently, lattice shaking has become a powerful
tool to manipulate ultracold atoms in optical lattices \cite{shaking1,shaking2,shaking3,shaking_rmp},
and it offers a direct access to the orbital DoF. Site-resolved lattice shakings have already been used
to design local quantum gates operating on the orbital DoF \cite{interaction_blockade}.
In this work, we propose a lattice shaking scheme, which can efficiently generate scalable lattice entanglement
encoded into the orbital DoF with a single operation.
The generated entangled states can be directly applied for MBQC, which
makes the lattice shaking scheme an elementary ingredient for MBQC with quantum information encoded in the orbital DoF.

\textit{Setup and preparation scheme.\textemdash}
We consider a one-dimensional (1D) double-well superlattice loaded with contact interacting bosons, with a unit filling per site,
$i.e.$ two bosons per unit cell. The Hamiltonian of the system is written as:
\begin{equation}\label{ham0}
 H_0=\sum_{i=1}^{2N}(\frac{-\hbar^2}{2M}\partial^2_{x_i}+V_{sl}(x_i))+\sum_{i<j=1}^{2N}g\delta(x_i,x_j),
\end{equation}
which describes $2N$ atoms of mass $M$ confined in a 1D double-well superlattice of $N$ unit cells, and the superlattice is given by
$V_{sl}(x)=V_0(\sin^2(kx/2)+2\cos^2(kx))$, which can be formed by two pairs of counter-propagating laser beams of wave vectors $k/2$ and $k$,
respectively.
In this work we consider a deep superlattice with $V_0=10\ E_R$, where $E_R\equiv\hbar^2k^2/2M$ is the recoil energy,
and each site possesses at least two well-defined
single-particle states, named as the s- and the p-orbital, respectively.

Employing a corresponding interaction strength, the ground state of the system is a Mott-like state,
in which each atom occupies a separate site, residing in the lowest orbital, $i.e.$ the s-orbital of the site.
This ground state can be described as:
\begin{equation}\label{G}
  |G\rangle=\prod_{i=1}^{N}|s,s\rangle_i,
\end{equation}
where $|\alpha,\beta\rangle_i$ denotes that two atoms occupy the $i$-th unit cell of the superlattice,
one in the $\alpha$-orbital of the left site, and the other in the $\beta$-orbital of the right site of the cell.
The targeted entangled states are chosen as:
\begin{equation}\label{pm}
  |\pm\rangle=\prod_{i=1}^{N}\frac{1}{\sqrt{2}}(|s,p\rangle_i\pm|p,s\rangle_i).
\end{equation}
Both $|+\rangle$ and $|-\rangle$ correspond to an entangled state, of which the two atoms in the same unit cell become entangled
with respect to their orbital DoF, with only one atom in the cell occupying the p-orbital and the other remaining in the s-orbital.
Residing in states $|\pm\rangle$, the whole system becomes a cluster of bipartite entangled pairs in each unit cell.
To a good approximation, $|\pm\rangle$ are degenerate eigenstates of $H_0$, and this ensures that
the system can stay stable in the entangled state after the generation scheme, which will be of benefit to further
operations on the entangled states, such as the quantum gate operations.

\begin{figure}
\includegraphics[width=15cm]{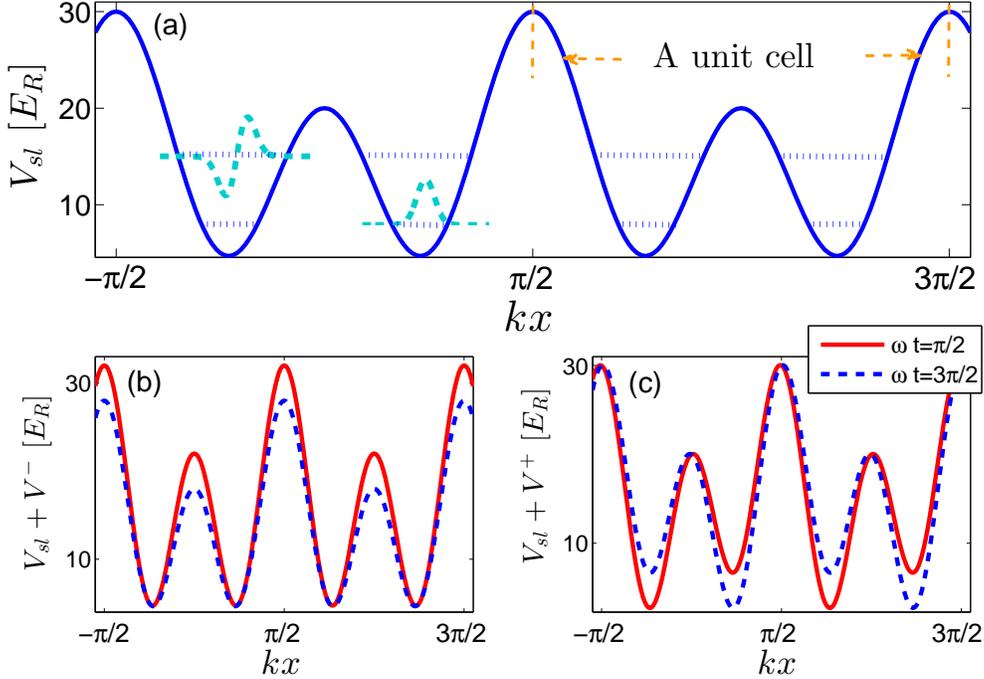}
\caption{ A sketch of the double-well superlattice (figure (a)), and the shaking potential of $V^-$ (figure (b)), and $V^+$ (figure (c)).
In the main figure, the energy levels and the profiles of the s- and p-orbitals are sketched by the blue doted and cyan dashed lines, respectively.
The potential of $V^-$ and $V^+$ with the largest shaking amplitudes are shown in figure (b) and (c), respectively,
where it can be seen that $V^-$ provides a shaking of the inter-site barriers,
whereas $V^+$ provides a temporal modulation of the energy offset between the left and right sites in a cell.} \label{fig1}
\end{figure}

Starting from the initial state $|G\rangle$, we propose a resonant shaking scheme to transfer the system to the targeted entangled states
$|+\rangle$ or $|-\rangle$, and the resonant shaking refers to a periodic modulation of the superlattice potential with the frequency
matching the energy difference between $|G\rangle$ and $|\pm\rangle$.
A prerequisite for the scheme is that the resonant shaking must be able to selectively excite one of
the degenerate eigenstates $|\pm\rangle$.
Otherwise, if both states are simultaneously excited, the entanglement would be diminished or even completely destroyed.
Two shaking potentials can realize the required selective excitation, which are
\begin{equation}\label{vsh1}
\begin{split}
V^+(x,t)&=V_+ \sin(\omega t)\sin(kx/2)\theta(t)\theta(\tau_+-t),\\
V^-(x,t)&=V_- \sin(\omega t)\cos^2(kx)\theta(t)\theta(\tau_--t).
\end{split}
\end{equation}
$V^{+(-)}$ can selectively excite the system from $|G\rangle$ to $|+\rangle$ ($|-\rangle$).
The product of step functions $\theta(t)\theta(\tau_\pm-t)$ indicates that the shaking is
instantaneously turned on and off at time $t=0$ and $t=\tau_\pm$. $\tau_\pm$ can be chosen
as the time when the shaking transfers the system to the targeted entangled state most completely.
Related shakings have been experimentally applied to realize controllable tunneling and coupling
in optical lattices \cite{lshaking2,lshaking1}.
The underlying mechanism of the selective excitation is a special two-body parity defined by a local transformation
applied to the pair of atoms in each unit cell. More specifically, let's
focus on a single unit cell, and consider the transformation of $\hat T: x_1\longleftrightarrow-x_2$, where $x_1$ and $x_2$ denote the two
atoms'
coordinates in the unit cell with the origin taken at the center of the cell.
It can be proven that $|G\rangle$ and $|\pm\rangle$ have a well defined parity upon this transformation with $\hat T|G\rangle=|G\rangle$
and $\hat T|\pm\rangle=\mp|\pm\rangle$. A shaking potential $V_s$ with $\hat TV_s\hat T^\dagger=(-)V_s$ will selectively couple the initial
state $|G\rangle$ to the entangled state $|-\rangle$ ($|+\rangle$).
It can be shown that $\hat T V^\pm \hat T^\dagger=\mp V^\pm$ , and a selective excitation of  $|+\rangle$ ($|-\rangle$)
can be achieved by the shaking potential $V^{+(-)}$.
The 1D double-well superlattice as well as the shaking potentials $V^\pm$ are
illustrated in figure \ref{fig1}.

To demonstrate the validity of the resonant shaking scheme, we perform $ab\ initio$ numerical simulations of the dynamical process of shaking.
The method applied here is the Multi-Layer Multi-Configuration Time-Dependent Hartree method
for Mixtures (ML-MCTDHX) \cite{mlb1,mlb2,mlx}.
This method is developed from MCTDH \cite{mctdh1,mctdh2}, and works equivalently to MCTDHB \cite{mctdhb1,mctdhb2,mctdhx},
when handling a single-species system of indistinguishable bosons. It accounts for all correlations among the bosons
and can intrinsically take
into account the exact geometry of the 1D superlattice and the parity of the shaking potentials, which turns out to be essential for the
generation scheme in this work. The atoms in the superlattice are taken as a closed system, and effects due to finite but sufficiently low
temperature and/or spontaneous emission are ignored, which, however,
we believe will not affect the main results of this work.

\textit{Validity and Accuracy.\textemdash}
We will now present evidence for the validity and accuracy of the lattice shaking scheme with
$V^\pm$ by performing $ab \ initio$ fully correlated simulations via ML-MCTDHX \cite{mlb1,mlb2,mlx}.
We hereby mainly focus on two aspects of the scheme: (i) Within a unit cell the lattice shaking can indeed
transfer the initial state to the targeted entangled state;
(ii) inter-cell interactions of neighboring cells will not affect the intra-cell entanglement.
The simulations are performed on a homogeneous double-well superlattice containing two unit cells,
where the intra- and inter-cell aspects concerning the validity of our scheme can be simultaneously addressed.
Periodic boundary conditions are used, which however do not affect the main results discussed here.
To characterize the outcome of the shaking process, two quantities are analyzed: The fidelity of the targeted entangled states
$f_\pm(t)=|\langle\pm|\Psi(t)\rangle|^2$ \cite{fidelity}, and the
two-body correlation function $g^{(2)}(x_1,x_2)=\rho_2(x_1,x_2)/(\rho_1(x_1)\rho_1(x_2))-1$, with $\rho_2(x_1,x_2)$
and $\rho_1(x_i)$
($i=1,2$) denoting the two-body and one-body densities, respectively.
The fidelity gives full access to the time evolution of the entangled states
during the shaking process, and the two-body correlation is used to analyze the generated entangled states.

Figure 2(a) and 2(b) present the results with shaking potential $V^+$ and $V^-$, respectively. The shaking is applied to the lattice for
a finite time period, and is turned off when the fidelity of the targeted entangled states reaches maximum. In figure 2(a),
during the shaking with $V^+$, the fidelity of $|+\rangle$ monotonically increases, reaching a maximum value around $95.4\%$,
which corresponds to a fidelity of $97.7\%$ in each unit cell.
When the shaking is turned off, the fidelity remains relatively stable, with a decay of $~0.8\%$ 
over a time of $100 \ \hbar/E_R$.
The temporal evolution of the fidelity indicates that the lattice shaking scheme with
$V^+$ can indeed excite the two atoms in a unit cell to the targeted entangled state $|+\rangle$, and maintain the system stable in the
entangled state after shaking, facilitating the proceeding operations on the entangled states, such as quantum gate operations.
The two-body correlation function provides a further indicator of the generation of the entangled states.
At the starting time (Fig. 2(a1)), the system resides in $|G\rangle$, and there is no nontrivial correlation
but the anti-bunching of the atoms in the diagonal blocks indicating the Mott-insulator like initial state
due to the repulsive contact interaction. At a later time when the shaking has
been turned off, the two atoms in the same unit cell get entangled,
documented by the wings appearing in the intra-cell correlation blocks.
Meanwhile, the inter-cell correlation remains almost zero in the complete dynamical process,
indicating that the inter-cell influence is vanishingly small on the intra-cell entanglement generation.
Similarly, figure 2(b) shows the fidelity and the two-body correlations as a function of time
under the lattice shaking of $V^-$. A maximum fidelity $f_-(t)$ of $94.0\%$ is reached,
corresponding to a fidelity of $97\%$ in each unit cell,
which decays by roughly $0.03\%$ within $70 \ \hbar/E_R$ after the shaking has been turned off.
In the two-body correlations, the expected wings in the intra-cell correlation blocks and
the vanishing inter-cell correlation are present.
In total, the fidelities and the two-body correlations confirm the validity of
the controllable entanglement generation scheme with a lattice shaking of $V^\pm$.

To realize a scalable cluster of bipartite entangled pairs in a superlattice with more unit cells,
a prerequisite is the homogeneity of the lattice.
In experiments, the overall harmonic confinement would lead to an inhomogeneous lattice
and vary the resonant shaking frequency between unit cells.
However, techniques to compensate for these confinement effects have been developed \cite{anticonfinement1,anticonfinement2},
which can be used to restore the homogeneity of the superlattice and ensure the scalability of the lattice shaking scheme.

\begin{figure}[htbp]
\centering
\subfloat{
\label{fig2a}
\includegraphics[width=13cm]{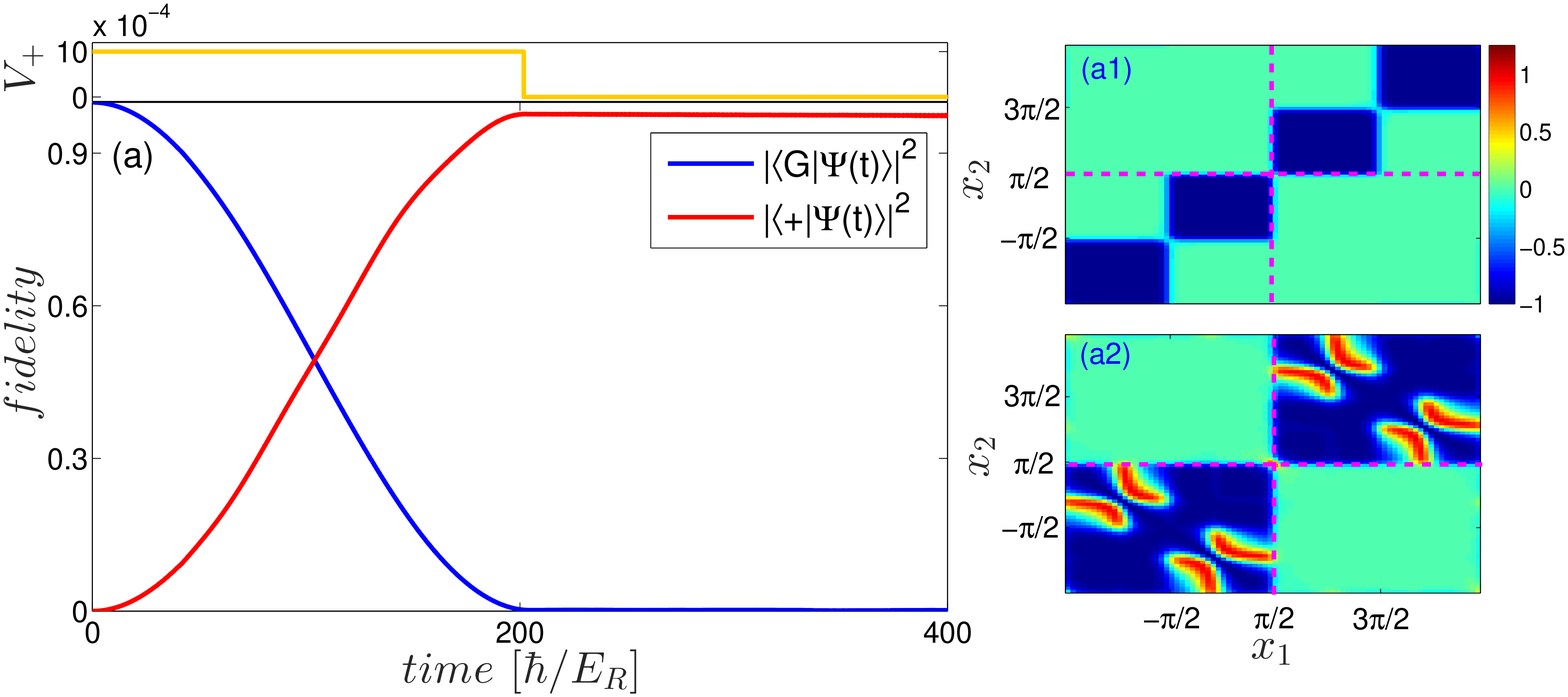}
}
\hfill
\subfloat{
\label{fig2b}
\includegraphics[width=13cm]{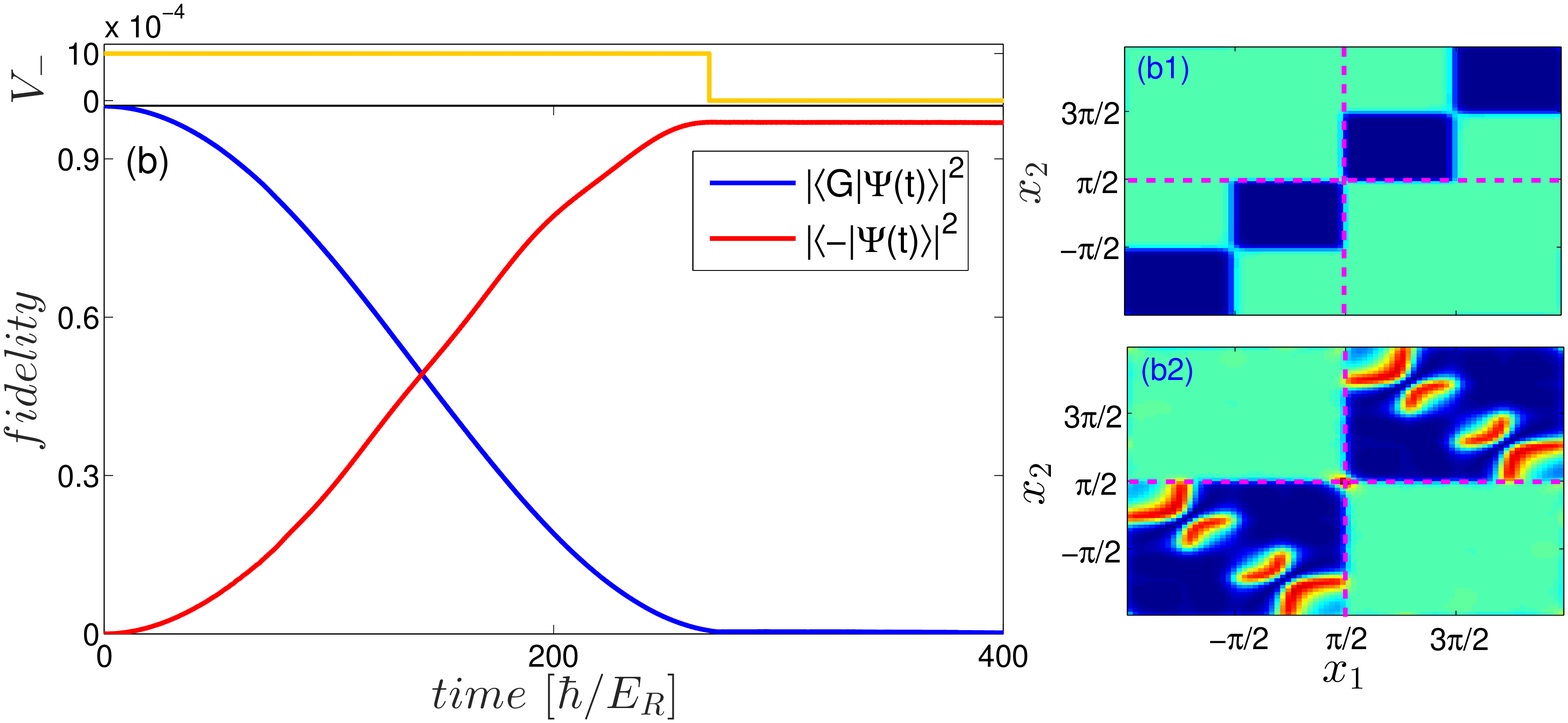}
}
\caption{ Figure (a) ((b)) present the fidelity of corresponding entangled state as a function of time,
under shaking via $V^+$ ($V^-$), and the upper panel shows the profile of the shaking amplitude,
with $V_{+(-)}=0.001\ E_R$ and $\omega=7.5475(7.544)\ E_R/\hbar$.
Figure (a1) and (b1) show the $g^{(2)}$ correlation function at the initial
time of the corresponding shaking, and (a2) and (b2) show it at a time $t=212\ \hbar/E_R$ and
$t=274\ \hbar/E_R$, respectively, when the shaking has been turned off. The four insets share the same color bar.}\label{fig2}
\end{figure}

\textit{Efficiency.\textemdash}
Two main characteristics of the efficiency of the entanglement generation are the maximum fidelity of the targeted state
and the preparation time to reach the maximum fidelity. In this section we investigate the dependence of
these two characteristics on the system parameters.
$Ab\ initio$ simulations (see previous section) confirm that the inter-cell influence is vanishingly small,
and the superlattice under shaking is effectively decoupled into a set of
double-well unit cells. It is then sufficient to perform simulations on a single unit cell to investigate the efficiency,
and the corresponding results are presented in this section.
The generation of $|-\rangle$ through shaking $V^-$ is taken as an example to demonstrate the dependence
on the relevant parameters.

The maximum fidelity is mainly determined by how many eigenstates other than the targeted entangled state
are populated during the shaking.
Indicated by the energy spectrum as a function of $g$, as shown in figure 3(a),
most of the co-excitations are shared by the eigenstates $|sp\rangle_-\equiv|(sp),0\rangle-|0,(sp)\rangle$,
$|ss\rangle_-\equiv|(ss),0\rangle-|0,(ss)\rangle$ and $|p,p\rangle$,
where $|(\alpha\beta),0\rangle$ ($|0,(\alpha\beta)\rangle$) denotes a double-occupation in the left (right)
site of the $\alpha$ and $\beta$ orbital.
$|sp\rangle_-$ ($|ss\rangle_-$) energetically lies close to $|-\rangle$ in the weak (strong) interaction regime,
and can be strongly excited by the shaking in the corresponding regime.
The energy difference between $|p,p\rangle$ and $|G\rangle$ is close to two times of that between $|-\rangle$ and $|G\rangle$,
and can be co-excited through a two-photon absorption process. The energy spectrum as shown in figure 3(a) illustrates
that the interaction strength controls the energetic decoupling of $|-\rangle$ from $|sp\rangle_-$, $|ss\rangle_-$ and $|p,p\rangle$,
$i.e.$ choosing the interaction strength in the intermediate regime can detune $|-\rangle$ 
from $|sp\rangle_-$ and $|ss\rangle_-$,
and meanwhile suppress the two-photon absorption by the orbital-dependent interaction strength of the
two atoms \cite{interaction_blockade}. Our simulation reveals that in the intermediate regime of $g$,
roughly speaking in the interval $g\in[1,5]$, the co-excitations of
unwanted eigenstates can be significantly suppressed, and the maximum fidelity is almost unchanged in this interval.
In the following,
we fix the interaction strength to $g=2.5$ and further investigate the effect of other relevant system parameters.

With a fixed interaction strength in the intermediate regime, the maximum fidelity and the preparation time also depend
on the shaking amplitude. Increasing the shaking amplitude would, on the one hand,
enhance the coupling between the initial and the target entangled state, which fastens the preparation time.
On the other hand, increasing the shaking amplitude would also lead to a stronger co-excitation of unwanted eigenstates
and reduce the maximum fidelity. This two-fold dependence is demonstrated in figure 3(b),
where both the maximum fidelity and the preparation time decrease as the shaking amplitude increases.
When the shaking amplitude increases, for instance, from $0.001$ to $0.005$,
the preparation time decreases almost by a factor of $5$, from $280\ \hbar/E_R$ to $60\ \hbar/E_R$.
Meanwhile the maximum fidelity just decreases from
$97.8\%$ to $95.4\%$, which indicates that this time depends more sensitively on the shaking amplitude than the
maximum fidelity. The maximum fidelity as a function of the frequency detuning from the resonant frequency is also an
important characteristic, due to the imperfect control of the shaking frequency in experiments.
Figure 3(c) shows such a dependence for different shaking amplitudes.
It is shown that for a stronger shaking amplitude, the maximum fidelity decreases more slowly with respect to
the frequency detuning. In use of the shaking scheme, it is necessary to find
a balance between the required maximum fidelity and preparation time, in order to optimize the shaking amplitude accordingly.

\begin{figure}[htbp]
\centering
\subfloat{
\label{fig3a}
\includegraphics[width=10cm]{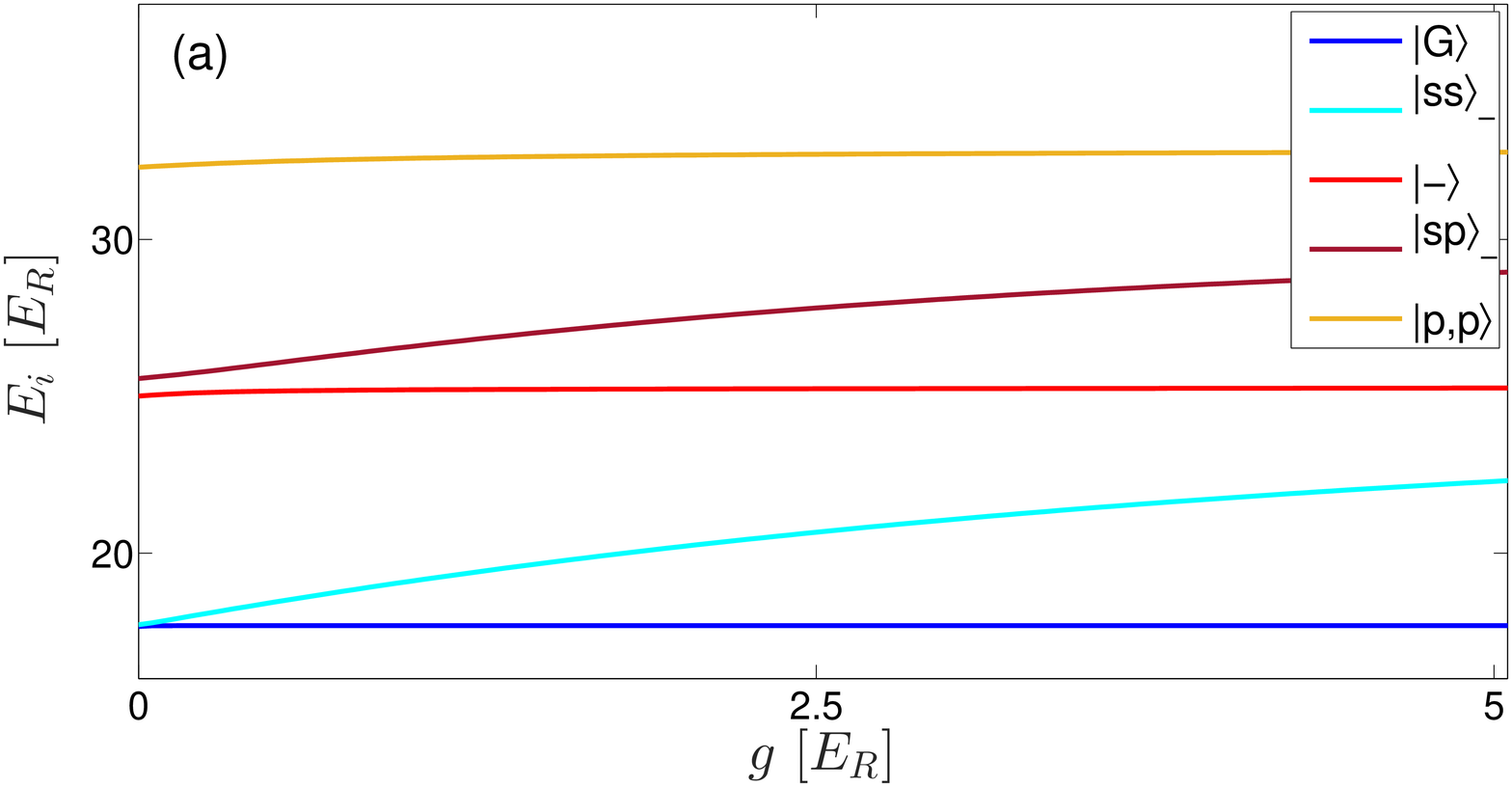}
}
\hfill
\subfloat{
\label{fig3b}
\includegraphics[width=10cm]{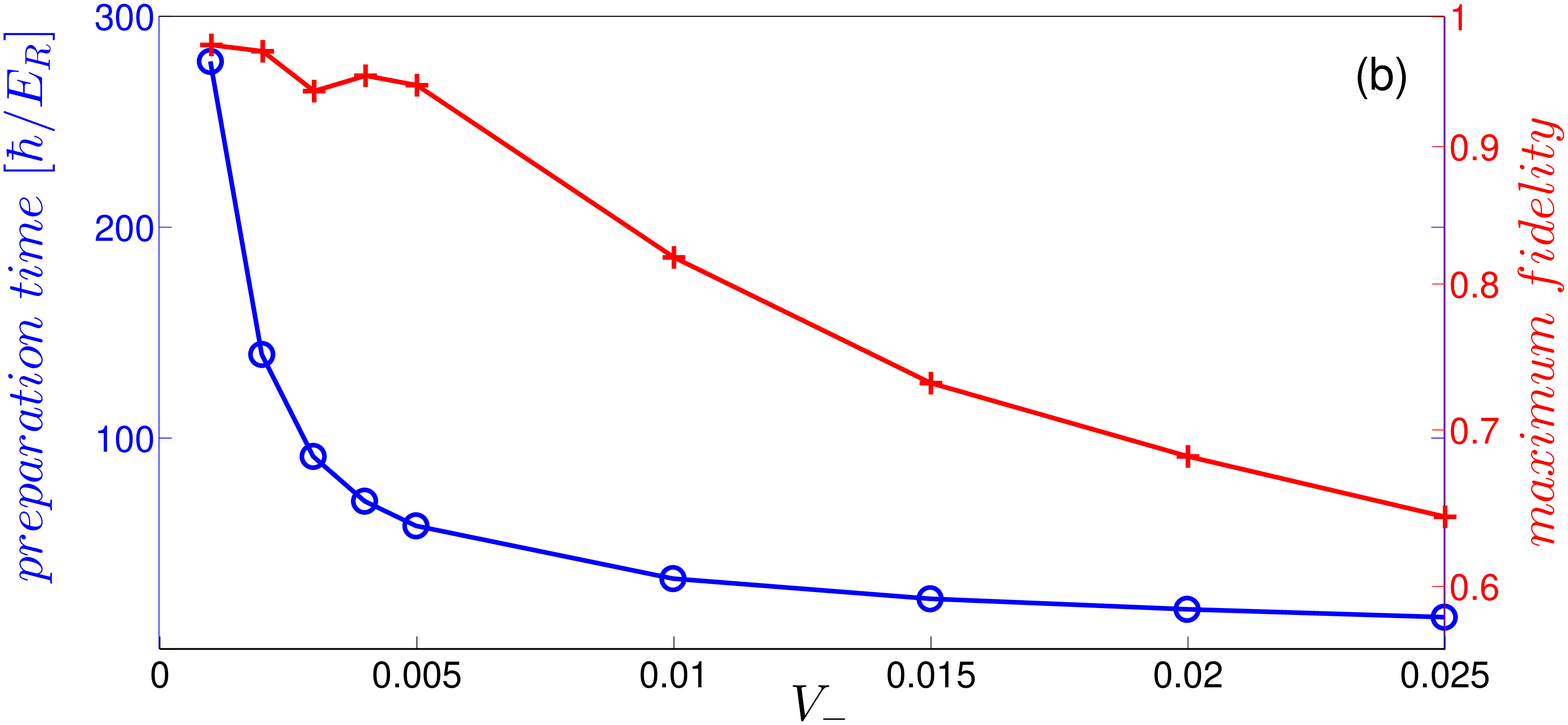}
}
\hfill
\subfloat{
\label{fig3c}
\includegraphics[width=10cm]{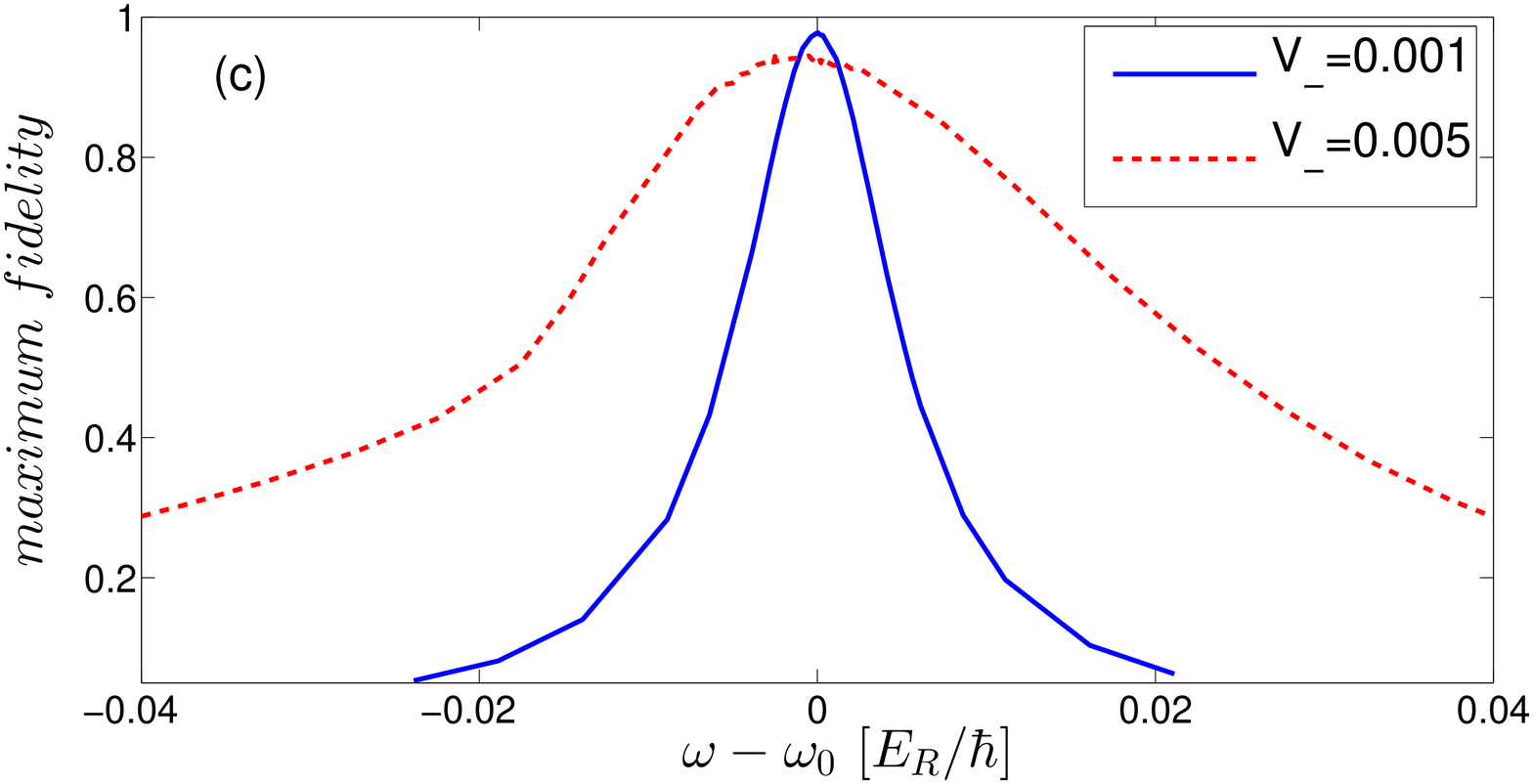}
}
\caption{(a) The eigenenergy spectrum $E_i$ of two bosons in a double well as a function of the interaction strength $g$,
where the eigenenergies of $|G\rangle$, $|-\rangle$, $|sp\rangle_-$, $|ss\rangle_-$ and $|p,p\rangle$ are shown.
(b) The maximum fidelity (right axis) and preparation time (left axis) as a function of the shaking amplitude $V_-$.
(c) The maximum fidelity as a function of the frequency detuning for a shaking amplitude $V_-=0.001$ (solid)
and $0.005$ (dashed).
}\label{fig3}
\end{figure}

\textit{Discussion and conclusions.\textemdash}
We have developed and investigated a lattice-shaking scheme to generate scalable entangled states encoded within the orbital
degrees of freedom of ultracold atoms in double-well superlattices.
This scheme can significantly simplify the procedure of generating a cluster of bipartite entangled states,
involving only a single step of resonant lattice shaking.
Considering an experimental accessible setup, $e.g.$, of $^{87}Rb$ atoms confined in a double-well superlattice
formed by lasers of wavelength $767\ nm$ and $1534\ nm$ \cite{spinDL} with $V_0=10\ E_R$,
the lattice shaking can generate desired entangled states of fidelity $>97\%$ with a preparation time less than $100\ \mu s$.
The fidelity can be further improved by increasing the lattice height and optimizing the shaking profile \cite{two-photon},
which will suppress the co-excitation of $|sp\rangle_\pm$ and $|ss\rangle_\pm$, as well as $|p,p\rangle$, respectively.
The generated entangled states $|\pm\rangle$ are a direct analogue to the resilient entangled state
proposed in Ref \cite{Jaksch}, with replacing the spin DoF by the orbital DoF.
Quantum gate operations for MBQC proposed in \cite{Jaksch} can be directly transfered to
our proposed states $|\pm\rangle$.
Moreover, several quantum gates have also been particularly designed for qubits encoded into
the orbital DoF \cite{porbital_2002,porbital_2002_2,porbital_2008,interaction_blockade},
and they are also readily applied to $|\pm\rangle$, which guarantees $|\pm\rangle$ as a promising resource for MBQC.
The lattice shaking scheme then contributes an essential ingredient to the MBQC based on the orbital DoF.
The lattice shaking scheme could also be applied to two-dimensional optical superlattices, where a true cluster
state of maximum entanglement can be generated through resonant drivings.

The orbital DoF also permits flexible manipulation and detection manners. For instance, direct lattice shaking can flip
the orbitals, resembling the spin flipping through Raman pulses. Although a direct detection of the orbital state is
difficult, the orbital state can be mapped to internal DoF or site occupations,
of which well developed techniques are ready for use.
For instance, using spin-dependent superlattices can realize site- and orbital-resolved spin flipping \cite{spinDL}
and map the orbital DoF to the internal DoF. Alternatively, one can map the s- and p-orbitals of
one site to the d- and f-orbitals of the other site, and a subsequent band mapping \cite{band-mapping1,band-mapping2,band-mapping3}
will count the particle numbers in different orbitals of the two sites.
These rich manipulation and detection schemes allow broader uses of the proposed entangled states besides the MBQC, such
as in the test of Bell's inequality \cite{dw_entanglement}.

\begin{acknowledgments}
This work is supported by the National Natural Science Foundation of China (Grant Nos. 91636221 and 11604107).
P. S. acknowledges financial support by the excellence cluster “The Hamburg Centre for Ultrafast Imaging
— Structure, Dynamics, and Control of Matter at the Atomic Scale ” of the Deutsche Forschungsgemeinschaft.

\end{acknowledgments}

\bibliographystyle{apsrev4-1}
\bibliography{Ref.bib}

\begin{thebibliography}{44}%
\makeatletter
\providecommand \@ifxundefined [1]{%
 \@ifx{#1\undefined}
}%
\providecommand \@ifnum [1]{%
 \ifnum #1\expandafter \@firstoftwo
 \else \expandafter \@secondoftwo
 \fi
}%
\providecommand \@ifx [1]{%
 \ifx #1\expandafter \@firstoftwo
 \else \expandafter \@secondoftwo
 \fi
}%
\providecommand \natexlab [1]{#1}%
\providecommand \enquote  [1]{``#1''}%
\providecommand \bibnamefont  [1]{#1}%
\providecommand \bibfnamefont [1]{#1}%
\providecommand \citenamefont [1]{#1}%
\providecommand \href@noop [0]{\@secondoftwo}%
\providecommand \href [0]{\begingroup \@sanitize@url \@href}%
\providecommand \@href[1]{\@@startlink{#1}\@@href}%
\providecommand \@@href[1]{\endgroup#1\@@endlink}%
\providecommand \@sanitize@url [0]{\catcode `\\12\catcode `\$12\catcode
  `\&12\catcode `\#12\catcode `\^12\catcode `\_12\catcode `\%12\relax}%
\providecommand \@@startlink[1]{}%
\providecommand \@@endlink[0]{}%
\providecommand \url  [0]{\begingroup\@sanitize@url \@url }%
\providecommand \@url [1]{\endgroup\@href {#1}{\urlprefix }}%
\providecommand \urlprefix  [0]{URL }%
\providecommand \Eprint [0]{\href }%
\providecommand \doibase [0]{http://dx.doi.org/}%
\providecommand \selectlanguage [0]{\@gobble}%
\providecommand \bibinfo  [0]{\@secondoftwo}%
\providecommand \bibfield  [0]{\@secondoftwo}%
\providecommand \translation [1]{[#1]}%
\providecommand \BibitemOpen [0]{}%
\providecommand \bibitemStop [0]{}%
\providecommand \bibitemNoStop [0]{.\EOS\space}%
\providecommand \EOS [0]{\spacefactor3000\relax}%
\providecommand \BibitemShut  [1]{\csname bibitem#1\endcsname}%
\let\auto@bib@innerbib\@empty
\bibitem [{\citenamefont {Bloch}(2005)}]{Bloch2005}%
  \BibitemOpen
  \bibfield  {author} {\bibinfo {author} {\bibfnamefont {I.}~\bibnamefont
  {Bloch}},\ }\href@noop {} {\bibfield  {journal} {\bibinfo  {journal} {Nat.
  Phys.}\ }\textbf {\bibinfo {volume} {1}},\ \bibinfo {pages} {23} (\bibinfo
  {year} {2005})}\BibitemShut {NoStop}%
\bibitem [{\citenamefont {Bloch}\ \emph {et~al.}(2008)\citenamefont {Bloch},
  \citenamefont {Dalibard},\ and\ \citenamefont {Zwerger}}]{Bloch2008c}%
  \BibitemOpen
  \bibfield  {author} {\bibinfo {author} {\bibfnamefont {I.}~\bibnamefont
  {Bloch}}, \bibinfo {author} {\bibfnamefont {J.}~\bibnamefont {Dalibard}}, \
  and\ \bibinfo {author} {\bibfnamefont {W.}~\bibnamefont {Zwerger}},\
  }\href@noop {} {\bibfield  {journal} {\bibinfo  {journal} {Rev. Mod. Phys.}\
  }\textbf {\bibinfo {volume} {80}},\ \bibinfo {pages} {885} (\bibinfo {year}
  {2008})}\BibitemShut {NoStop}%
\bibitem [{\citenamefont {Luo}\ \emph {et~al.}(2017)\citenamefont {Luo},
  \citenamefont {Zou}, \citenamefont {Wu}, \citenamefont {Liu}, \citenamefont
  {Han}, \citenamefont {Tey},\ and\ \citenamefont
  {You}}]{entanglement_measurement}%
  \BibitemOpen
  \bibfield  {author} {\bibinfo {author} {\bibfnamefont {X.-Y.}\ \bibnamefont
  {Luo}}, \bibinfo {author} {\bibfnamefont {Y.-Q.}\ \bibnamefont {Zou}},
  \bibinfo {author} {\bibfnamefont {L.-N.}\ \bibnamefont {Wu}}, \bibinfo
  {author} {\bibfnamefont {Q.}~\bibnamefont {Liu}}, \bibinfo {author}
  {\bibfnamefont {M.-F.}\ \bibnamefont {Han}}, \bibinfo {author} {\bibfnamefont
  {M.~K.}\ \bibnamefont {Tey}}, \ and\ \bibinfo {author} {\bibfnamefont
  {L.}~\bibnamefont {You}},\ }\href@noop {} {\bibfield  {journal} {\bibinfo
  {journal} {Science}\ }\textbf {\bibinfo {volume} {355}},\ \bibinfo {pages}
  {620} (\bibinfo {year} {2017})}\BibitemShut {NoStop}%
\bibitem [{\citenamefont {Jaksch}\ \emph {et~al.}(1999)\citenamefont {Jaksch},
  \citenamefont {Briegel}, \citenamefont {Cirac}, \citenamefont {Gardiner},\
  and\ \citenamefont {Zoller}}]{original_QC}%
  \BibitemOpen
  \bibfield  {author} {\bibinfo {author} {\bibfnamefont {D.}~\bibnamefont
  {Jaksch}}, \bibinfo {author} {\bibfnamefont {H.-J.}\ \bibnamefont {Briegel}},
  \bibinfo {author} {\bibfnamefont {J.~I.}\ \bibnamefont {Cirac}}, \bibinfo
  {author} {\bibfnamefont {C.~W.}\ \bibnamefont {Gardiner}}, \ and\ \bibinfo
  {author} {\bibfnamefont {P.}~\bibnamefont {Zoller}},\ }\href@noop {}
  {\bibfield  {journal} {\bibinfo  {journal} {Phys. Rev. Lett.}\ }\textbf
  {\bibinfo {volume} {82}},\ \bibinfo {pages} {1975} (\bibinfo {year}
  {1999})}\BibitemShut {NoStop}%
\bibitem [{\citenamefont {Brennen}\ \emph {et~al.}(1999)\citenamefont
  {Brennen}, \citenamefont {Caves}, \citenamefont {Jessen},\ and\ \citenamefont
  {Deutsch}}]{original_QC2}%
  \BibitemOpen
  \bibfield  {author} {\bibinfo {author} {\bibfnamefont {G.~K.}\ \bibnamefont
  {Brennen}}, \bibinfo {author} {\bibfnamefont {C.~M.}\ \bibnamefont {Caves}},
  \bibinfo {author} {\bibfnamefont {P.~S.}\ \bibnamefont {Jessen}}, \ and\
  \bibinfo {author} {\bibfnamefont {I.~H.}\ \bibnamefont {Deutsch}},\
  }\href@noop {} {\bibfield  {journal} {\bibinfo  {journal} {Phys. Rev. Lett.}\
  }\textbf {\bibinfo {volume} {82}},\ \bibinfo {pages} {1060} (\bibinfo {year}
  {1999})}\BibitemShut {NoStop}%
\bibitem [{\citenamefont {Briegel}\ \emph {et~al.}(2009)\citenamefont
  {Briegel}, \citenamefont {Browne}, \citenamefont {Dur}, \citenamefont
  {Raussendorf},\ and\ \citenamefont {Van~den Nest}}]{MBQC}%
  \BibitemOpen
  \bibfield  {author} {\bibinfo {author} {\bibfnamefont {H.~J.}\ \bibnamefont
  {Briegel}}, \bibinfo {author} {\bibfnamefont {D.~E.}\ \bibnamefont {Browne}},
  \bibinfo {author} {\bibfnamefont {W.}~\bibnamefont {Dur}}, \bibinfo {author}
  {\bibfnamefont {R.}~\bibnamefont {Raussendorf}}, \ and\ \bibinfo {author}
  {\bibfnamefont {M.}~\bibnamefont {Van~den Nest}},\ }\href@noop {} {\bibfield
  {journal} {\bibinfo  {journal} {Nat. Phys.}\ }\textbf {\bibinfo {volume}
  {5}},\ \bibinfo {pages} {19} (\bibinfo {year} {2009})}\BibitemShut {NoStop}%
\bibitem [{\citenamefont {Briegel}\ and\ \citenamefont
  {Raussendorf}(2001)}]{cluster_state}%
  \BibitemOpen
  \bibfield  {author} {\bibinfo {author} {\bibfnamefont {H.~J.}\ \bibnamefont
  {Briegel}}\ and\ \bibinfo {author} {\bibfnamefont {R.}~\bibnamefont
  {Raussendorf}},\ }\href@noop {} {\bibfield  {journal} {\bibinfo  {journal}
  {Phys. Rev. Lett.}\ }\textbf {\bibinfo {volume} {86}},\ \bibinfo {pages}
  {910} (\bibinfo {year} {2001})}\BibitemShut {NoStop}%
\bibitem [{\citenamefont {Jiang}\ \emph {et~al.}(2009)\citenamefont {Jiang},
  \citenamefont {Rey}, \citenamefont {Romero-Isart}, \citenamefont
  {Garc\'{\i}a-Ripoll}, \citenamefont {Sanpera},\ and\ \citenamefont
  {Lukin}}]{cluster}%
  \BibitemOpen
  \bibfield  {author} {\bibinfo {author} {\bibfnamefont {L.}~\bibnamefont
  {Jiang}}, \bibinfo {author} {\bibfnamefont {A.~M.}\ \bibnamefont {Rey}},
  \bibinfo {author} {\bibfnamefont {O.}~\bibnamefont {Romero-Isart}}, \bibinfo
  {author} {\bibfnamefont {J.~J.}\ \bibnamefont {Garc\'{\i}a-Ripoll}}, \bibinfo
  {author} {\bibfnamefont {A.}~\bibnamefont {Sanpera}}, \ and\ \bibinfo
  {author} {\bibfnamefont {M.~D.}\ \bibnamefont {Lukin}},\ }\href@noop {}
  {\bibfield  {journal} {\bibinfo  {journal} {Phys. Rev. A}\ }\textbf {\bibinfo
  {volume} {79}},\ \bibinfo {pages} {022309} (\bibinfo {year}
  {2009})}\BibitemShut {NoStop}%
\bibitem [{\citenamefont {Calarco}\ \emph {et~al.}(2000)\citenamefont
  {Calarco}, \citenamefont {Hinds}, \citenamefont {Jaksch}, \citenamefont
  {Schmiedmayer}, \citenamefont {Cirac},\ and\ \citenamefont
  {Zoller}}]{ccollision_original}%
  \BibitemOpen
  \bibfield  {author} {\bibinfo {author} {\bibfnamefont {T.}~\bibnamefont
  {Calarco}}, \bibinfo {author} {\bibfnamefont {E.~A.}\ \bibnamefont {Hinds}},
  \bibinfo {author} {\bibfnamefont {D.}~\bibnamefont {Jaksch}}, \bibinfo
  {author} {\bibfnamefont {J.}~\bibnamefont {Schmiedmayer}}, \bibinfo {author}
  {\bibfnamefont {J.~I.}\ \bibnamefont {Cirac}}, \ and\ \bibinfo {author}
  {\bibfnamefont {P.}~\bibnamefont {Zoller}},\ }\href@noop {} {\bibfield
  {journal} {\bibinfo  {journal} {Phys. Rev. A}\ }\textbf {\bibinfo {volume}
  {61}},\ \bibinfo {pages} {022304} (\bibinfo {year} {2000})}\BibitemShut
  {NoStop}%
\bibitem [{\citenamefont {Mandel}\ \emph {et~al.}(2003)\citenamefont {Mandel},
  \citenamefont {Greiner}, \citenamefont {Widera}, \citenamefont {Rom},
  \citenamefont {Hänsch},\ and\ \citenamefont {Bloch}}]{controlled_collision}%
  \BibitemOpen
  \bibfield  {author} {\bibinfo {author} {\bibfnamefont {O.}~\bibnamefont
  {Mandel}}, \bibinfo {author} {\bibfnamefont {M.}~\bibnamefont {Greiner}},
  \bibinfo {author} {\bibfnamefont {A.}~\bibnamefont {Widera}}, \bibinfo
  {author} {\bibfnamefont {T.}~\bibnamefont {Rom}}, \bibinfo {author}
  {\bibfnamefont {T.~W.}\ \bibnamefont {Hänsch}}, \ and\ \bibinfo {author}
  {\bibfnamefont {I.}~\bibnamefont {Bloch}},\ }\href@noop {} {\bibfield
  {journal} {\bibinfo  {journal} {Nature}\ }\textbf {\bibinfo {volume} {425}},\
  \bibinfo {pages} {937} (\bibinfo {year} {2003})}\BibitemShut {NoStop}%
\bibitem [{\citenamefont {Duan}\ \emph {et~al.}(2003)\citenamefont {Duan},
  \citenamefont {Demler},\ and\ \citenamefont
  {Lukin}}]{superexchange_original}%
  \BibitemOpen
  \bibfield  {author} {\bibinfo {author} {\bibfnamefont {L.-M.}\ \bibnamefont
  {Duan}}, \bibinfo {author} {\bibfnamefont {E.}~\bibnamefont {Demler}}, \ and\
  \bibinfo {author} {\bibfnamefont {M.~D.}\ \bibnamefont {Lukin}},\ }\href@noop
  {} {\bibfield  {journal} {\bibinfo  {journal} {Phys. Rev. Lett.}\ }\textbf
  {\bibinfo {volume} {91}},\ \bibinfo {pages} {090402} (\bibinfo {year}
  {2003})}\BibitemShut {NoStop}%
\bibitem [{\citenamefont {Anderlini}\ \emph {et~al.}(2007)\citenamefont
  {Anderlini}, \citenamefont {Lee}, \citenamefont {Brown}, \citenamefont
  {Sebby-Strabley}, \citenamefont {Phillips},\ and\ \citenamefont
  {Porto}}]{ccollision}%
  \BibitemOpen
  \bibfield  {author} {\bibinfo {author} {\bibfnamefont {M.}~\bibnamefont
  {Anderlini}}, \bibinfo {author} {\bibfnamefont {P.~J.}\ \bibnamefont {Lee}},
  \bibinfo {author} {\bibfnamefont {B.~L.}\ \bibnamefont {Brown}}, \bibinfo
  {author} {\bibfnamefont {J.}~\bibnamefont {Sebby-Strabley}}, \bibinfo
  {author} {\bibfnamefont {W.~D.}\ \bibnamefont {Phillips}}, \ and\ \bibinfo
  {author} {\bibfnamefont {J.~V.}\ \bibnamefont {Porto}},\ }\href@noop {}
  {\bibfield  {journal} {\bibinfo  {journal} {Nature}\ }\textbf {\bibinfo
  {volume} {448}},\ \bibinfo {pages} {452} (\bibinfo {year}
  {2007})}\BibitemShut {NoStop}%
\bibitem [{\citenamefont {Trotzky}\ \emph {et~al.}(2008)\citenamefont
  {Trotzky}, \citenamefont {Cheinet}, \citenamefont {F{\"o}lling},
  \citenamefont {Feld}, \citenamefont {Schnorrberger}, \citenamefont {Rey},
  \citenamefont {Polkovnikov}, \citenamefont {Demler}, \citenamefont {Lukin},\
  and\ \citenamefont {Bloch}}]{superexchange}%
  \BibitemOpen
  \bibfield  {author} {\bibinfo {author} {\bibfnamefont {S.}~\bibnamefont
  {Trotzky}}, \bibinfo {author} {\bibfnamefont {P.}~\bibnamefont {Cheinet}},
  \bibinfo {author} {\bibfnamefont {S.}~\bibnamefont {F{\"o}lling}}, \bibinfo
  {author} {\bibfnamefont {M.}~\bibnamefont {Feld}}, \bibinfo {author}
  {\bibfnamefont {U.}~\bibnamefont {Schnorrberger}}, \bibinfo {author}
  {\bibfnamefont {A.~M.}\ \bibnamefont {Rey}}, \bibinfo {author} {\bibfnamefont
  {A.}~\bibnamefont {Polkovnikov}}, \bibinfo {author} {\bibfnamefont {E.~A.}\
  \bibnamefont {Demler}}, \bibinfo {author} {\bibfnamefont {M.~D.}\
  \bibnamefont {Lukin}}, \ and\ \bibinfo {author} {\bibfnamefont
  {I.}~\bibnamefont {Bloch}},\ }\href@noop {} {\bibfield  {journal} {\bibinfo
  {journal} {Science}\ }\textbf {\bibinfo {volume} {319}},\ \bibinfo {pages}
  {295} (\bibinfo {year} {2008})}\BibitemShut {NoStop}%
\bibitem [{\citenamefont {Vaucher}\ \emph {et~al.}(2008)\citenamefont
  {Vaucher}, \citenamefont {Nunnenkamp},\ and\ \citenamefont
  {Jaksch}}]{Jaksch}%
  \BibitemOpen
  \bibfield  {author} {\bibinfo {author} {\bibfnamefont {B.}~\bibnamefont
  {Vaucher}}, \bibinfo {author} {\bibfnamefont {A.}~\bibnamefont {Nunnenkamp}},
  \ and\ \bibinfo {author} {\bibfnamefont {D.}~\bibnamefont {Jaksch}},\
  }\href@noop {} {\bibfield  {journal} {\bibinfo  {journal} {New J. Phys.}\
  }\textbf {\bibinfo {volume} {10}},\ \bibinfo {pages} {023005} (\bibinfo
  {year} {2008})}\BibitemShut {NoStop}%
\bibitem [{\citenamefont {Kaufman}\ \emph {et~al.}(2015)\citenamefont
  {Kaufman}, \citenamefont {Lester}, \citenamefont {Foss-Feig}, \citenamefont
  {Wall}, \citenamefont {Rey},\ and\ \citenamefont {Regal}}]{tweezer}%
  \BibitemOpen
  \bibfield  {author} {\bibinfo {author} {\bibfnamefont {A.~M.}\ \bibnamefont
  {Kaufman}}, \bibinfo {author} {\bibfnamefont {B.~J.}\ \bibnamefont {Lester}},
  \bibinfo {author} {\bibfnamefont {M.}~\bibnamefont {Foss-Feig}}, \bibinfo
  {author} {\bibfnamefont {M.~L.}\ \bibnamefont {Wall}}, \bibinfo {author}
  {\bibfnamefont {A.~M.}\ \bibnamefont {Rey}}, \ and\ \bibinfo {author}
  {\bibfnamefont {C.~A.}\ \bibnamefont {Regal}},\ }\href@noop {} {\bibfield
  {journal} {\bibinfo  {journal} {Nature}\ }\textbf {\bibinfo {volume} {527}},\
  \bibinfo {pages} {208} (\bibinfo {year} {2015})}\BibitemShut {NoStop}%
\bibitem [{\citenamefont {Dai}\ \emph {et~al.}(2016)\citenamefont {Dai},
  \citenamefont {Yang}, \citenamefont {Reingruber}, \citenamefont {Xu},
  \citenamefont {Jiang}, \citenamefont {Chen}, \citenamefont {Yuan},\ and\
  \citenamefont {Pan}}]{dw_entanglement}%
  \BibitemOpen
  \bibfield  {author} {\bibinfo {author} {\bibfnamefont {H.-N.}\ \bibnamefont
  {Dai}}, \bibinfo {author} {\bibfnamefont {B.}~\bibnamefont {Yang}}, \bibinfo
  {author} {\bibfnamefont {A.}~\bibnamefont {Reingruber}}, \bibinfo {author}
  {\bibfnamefont {X.-F.}\ \bibnamefont {Xu}}, \bibinfo {author} {\bibfnamefont
  {X.}~\bibnamefont {Jiang}}, \bibinfo {author} {\bibfnamefont {Y.-A.}\
  \bibnamefont {Chen}}, \bibinfo {author} {\bibfnamefont {Z.-S.}\ \bibnamefont
  {Yuan}}, \ and\ \bibinfo {author} {\bibfnamefont {J.-W.}\ \bibnamefont
  {Pan}},\ }\href@noop {} {\bibfield  {journal} {\bibinfo  {journal} {Nat.
  Phys.}\ }\textbf {\bibinfo {volume} {12}},\ \bibinfo {pages} {783} (\bibinfo
  {year} {2016})}\BibitemShut {NoStop}%
\bibitem [{\citenamefont {Li}\ and\ \citenamefont
  {Liu}(2016)}]{porbital_review}%
  \BibitemOpen
  \bibfield  {author} {\bibinfo {author} {\bibfnamefont {X.}~\bibnamefont
  {Li}}\ and\ \bibinfo {author} {\bibfnamefont {W.~V.}\ \bibnamefont {Liu}},\
  }\href@noop {} {\bibfield  {journal} {\bibinfo  {journal} {Reports on
  Progress in Physics}\ }\textbf {\bibinfo {volume} {79}},\ \bibinfo {pages}
  {116401} (\bibinfo {year} {2016})}\BibitemShut {NoStop}%
\bibitem [{\citenamefont {Charron}\ \emph {et~al.}(2002)\citenamefont
  {Charron}, \citenamefont {Tiesinga}, \citenamefont {Mies},\ and\
  \citenamefont {Williams}}]{porbital_2002}%
  \BibitemOpen
  \bibfield  {author} {\bibinfo {author} {\bibfnamefont {E.}~\bibnamefont
  {Charron}}, \bibinfo {author} {\bibfnamefont {E.}~\bibnamefont {Tiesinga}},
  \bibinfo {author} {\bibfnamefont {F.}~\bibnamefont {Mies}}, \ and\ \bibinfo
  {author} {\bibfnamefont {C.}~\bibnamefont {Williams}},\ }\href@noop {}
  {\bibfield  {journal} {\bibinfo  {journal} {Phys. Rev. Lett.}\ }\textbf
  {\bibinfo {volume} {88}},\ \bibinfo {pages} {077901} (\bibinfo {year}
  {2002})}\BibitemShut {NoStop}%
\bibitem [{\citenamefont {Eckert}\ \emph {et~al.}(2002)\citenamefont {Eckert},
  \citenamefont {Mompart}, \citenamefont {Yi}, \citenamefont {Schliemann},
  \citenamefont {Bru\ss{}}, \citenamefont {Birkl},\ and\ \citenamefont
  {Lewenstein}}]{porbital_2002_2}%
  \BibitemOpen
  \bibfield  {author} {\bibinfo {author} {\bibfnamefont {K.}~\bibnamefont
  {Eckert}}, \bibinfo {author} {\bibfnamefont {J.}~\bibnamefont {Mompart}},
  \bibinfo {author} {\bibfnamefont {X.~X.}\ \bibnamefont {Yi}}, \bibinfo
  {author} {\bibfnamefont {J.}~\bibnamefont {Schliemann}}, \bibinfo {author}
  {\bibfnamefont {D.}~\bibnamefont {Bru\ss{}}}, \bibinfo {author}
  {\bibfnamefont {G.}~\bibnamefont {Birkl}}, \ and\ \bibinfo {author}
  {\bibfnamefont {M.}~\bibnamefont {Lewenstein}},\ }\href@noop {} {\bibfield
  {journal} {\bibinfo  {journal} {Phys. Rev. A}\ }\textbf {\bibinfo {volume}
  {66}},\ \bibinfo {pages} {042317} (\bibinfo {year} {2002})}\BibitemShut
  {NoStop}%
\bibitem [{\citenamefont {Strauch}\ \emph {et~al.}(2008)\citenamefont
  {Strauch}, \citenamefont {Edwards}, \citenamefont {Tiesinga}, \citenamefont
  {Williams},\ and\ \citenamefont {Clark}}]{porbital_2008}%
  \BibitemOpen
  \bibfield  {author} {\bibinfo {author} {\bibfnamefont {F.~W.}\ \bibnamefont
  {Strauch}}, \bibinfo {author} {\bibfnamefont {M.}~\bibnamefont {Edwards}},
  \bibinfo {author} {\bibfnamefont {E.}~\bibnamefont {Tiesinga}}, \bibinfo
  {author} {\bibfnamefont {C.}~\bibnamefont {Williams}}, \ and\ \bibinfo
  {author} {\bibfnamefont {C.~W.}\ \bibnamefont {Clark}},\ }\href@noop {}
  {\bibfield  {journal} {\bibinfo  {journal} {Phys. Rev. A}\ }\textbf {\bibinfo
  {volume} {77}},\ \bibinfo {pages} {050304} (\bibinfo {year}
  {2008})}\BibitemShut {NoStop}%
\bibitem [{\citenamefont {Schneider}\ and\ \citenamefont
  {Saenz}(2012)}]{interaction_blockade}%
  \BibitemOpen
  \bibfield  {author} {\bibinfo {author} {\bibfnamefont {P.-I.}\ \bibnamefont
  {Schneider}}\ and\ \bibinfo {author} {\bibfnamefont {A.}~\bibnamefont
  {Saenz}},\ }\href@noop {} {\bibfield  {journal} {\bibinfo  {journal} {Phys.
  Rev. A}\ }\textbf {\bibinfo {volume} {85}},\ \bibinfo {pages} {050304}
  (\bibinfo {year} {2012})}\BibitemShut {NoStop}%
\bibitem [{\citenamefont {Inaba}\ \emph {et~al.}(2014)\citenamefont {Inaba},
  \citenamefont {Tokunaga}, \citenamefont {Tamaki}, \citenamefont {Igeta},\
  and\ \citenamefont {Yamashita}}]{porbital_interm}%
  \BibitemOpen
  \bibfield  {author} {\bibinfo {author} {\bibfnamefont {K.}~\bibnamefont
  {Inaba}}, \bibinfo {author} {\bibfnamefont {Y.}~\bibnamefont {Tokunaga}},
  \bibinfo {author} {\bibfnamefont {K.}~\bibnamefont {Tamaki}}, \bibinfo
  {author} {\bibfnamefont {K.}~\bibnamefont {Igeta}}, \ and\ \bibinfo {author}
  {\bibfnamefont {M.}~\bibnamefont {Yamashita}},\ }\href@noop {} {\bibfield
  {journal} {\bibinfo  {journal} {Phys. Rev. Lett.}\ }\textbf {\bibinfo
  {volume} {112}},\ \bibinfo {pages} {110501} (\bibinfo {year}
  {2014})}\BibitemShut {NoStop}%
\bibitem [{\citenamefont {Struck}\ \emph {et~al.}(2012)\citenamefont {Struck},
  \citenamefont {\"Olschl\"ager}, \citenamefont {Weinberg}, \citenamefont
  {Hauke}, \citenamefont {Simonet}, \citenamefont {Eckardt}, \citenamefont
  {Lewenstein}, \citenamefont {Sengstock},\ and\ \citenamefont
  {Windpassinger}}]{shaking1}%
  \BibitemOpen
  \bibfield  {author} {\bibinfo {author} {\bibfnamefont {J.}~\bibnamefont
  {Struck}}, \bibinfo {author} {\bibfnamefont {C.}~\bibnamefont
  {\"Olschl\"ager}}, \bibinfo {author} {\bibfnamefont {M.}~\bibnamefont
  {Weinberg}}, \bibinfo {author} {\bibfnamefont {P.}~\bibnamefont {Hauke}},
  \bibinfo {author} {\bibfnamefont {J.}~\bibnamefont {Simonet}}, \bibinfo
  {author} {\bibfnamefont {A.}~\bibnamefont {Eckardt}}, \bibinfo {author}
  {\bibfnamefont {M.}~\bibnamefont {Lewenstein}}, \bibinfo {author}
  {\bibfnamefont {K.}~\bibnamefont {Sengstock}}, \ and\ \bibinfo {author}
  {\bibfnamefont {P.}~\bibnamefont {Windpassinger}},\ }\href@noop {} {\bibfield
   {journal} {\bibinfo  {journal} {Phys. Rev. Lett.}\ }\textbf {\bibinfo
  {volume} {108}},\ \bibinfo {pages} {225304} (\bibinfo {year}
  {2012})}\BibitemShut {NoStop}%
\bibitem [{\citenamefont {Fläschner}\ \emph {et~al.}(2016)\citenamefont
  {Fläschner}, \citenamefont {Rem}, \citenamefont {Tarnowski}, \citenamefont
  {Vogel}, \citenamefont {Lühmann}, \citenamefont {Sengstock},\ and\
  \citenamefont {Weitenberg}}]{shaking2}%
  \BibitemOpen
  \bibfield  {author} {\bibinfo {author} {\bibfnamefont {N.}~\bibnamefont
  {Fläschner}}, \bibinfo {author} {\bibfnamefont {B.~S.}\ \bibnamefont {Rem}},
  \bibinfo {author} {\bibfnamefont {M.}~\bibnamefont {Tarnowski}}, \bibinfo
  {author} {\bibfnamefont {D.}~\bibnamefont {Vogel}}, \bibinfo {author}
  {\bibfnamefont {D.-S.}\ \bibnamefont {Lühmann}}, \bibinfo {author}
  {\bibfnamefont {K.}~\bibnamefont {Sengstock}}, \ and\ \bibinfo {author}
  {\bibfnamefont {C.}~\bibnamefont {Weitenberg}},\ }\href@noop {} {\bibfield
  {journal} {\bibinfo  {journal} {Science}\ }\textbf {\bibinfo {volume}
  {352}},\ \bibinfo {pages} {1091} (\bibinfo {year} {2016})}\BibitemShut
  {NoStop}%
\bibitem [{\citenamefont {Goldman}\ and\ \citenamefont
  {Dalibard}(2014)}]{shaking3}%
  \BibitemOpen
  \bibfield  {author} {\bibinfo {author} {\bibfnamefont {N.}~\bibnamefont
  {Goldman}}\ and\ \bibinfo {author} {\bibfnamefont {J.}~\bibnamefont
  {Dalibard}},\ }\href@noop {} {\bibfield  {journal} {\bibinfo  {journal}
  {Phys. Rev. X}\ }\textbf {\bibinfo {volume} {4}},\ \bibinfo {pages} {031027}
  (\bibinfo {year} {2014})}\BibitemShut {NoStop}%
\bibitem [{\citenamefont {Eckardt}(2017)}]{shaking_rmp}%
  \BibitemOpen
  \bibfield  {author} {\bibinfo {author} {\bibfnamefont {A.}~\bibnamefont
  {Eckardt}},\ }\href@noop {} {\bibfield  {journal} {\bibinfo  {journal} {Rev.
  Mod. Phys.}\ }\textbf {\bibinfo {volume} {89}},\ \bibinfo {pages} {011004}
  (\bibinfo {year} {2017})}\BibitemShut {NoStop}%
\bibitem [{\citenamefont {Trotzky}\ \emph {et~al.}(2010)\citenamefont
  {Trotzky}, \citenamefont {Chen}, \citenamefont {Schnorrberger}, \citenamefont
  {Cheinet},\ and\ \citenamefont {Bloch}}]{lshaking2}%
  \BibitemOpen
  \bibfield  {author} {\bibinfo {author} {\bibfnamefont {S.}~\bibnamefont
  {Trotzky}}, \bibinfo {author} {\bibfnamefont {Y.-A.}\ \bibnamefont {Chen}},
  \bibinfo {author} {\bibfnamefont {U.}~\bibnamefont {Schnorrberger}}, \bibinfo
  {author} {\bibfnamefont {P.}~\bibnamefont {Cheinet}}, \ and\ \bibinfo
  {author} {\bibfnamefont {I.}~\bibnamefont {Bloch}},\ }\href {\doibase
  10.1103/PhysRevLett.105.265303} {\bibfield  {journal} {\bibinfo  {journal}
  {Phys. Rev. Lett.}\ }\textbf {\bibinfo {volume} {105}},\ \bibinfo {pages}
  {265303} (\bibinfo {year} {2010})}\BibitemShut {NoStop}%
\bibitem [{\citenamefont {Chen}\ \emph {et~al.}(2011)\citenamefont {Chen},
  \citenamefont {Nascimb\`ene}, \citenamefont {Aidelsburger}, \citenamefont
  {Atala}, \citenamefont {Trotzky},\ and\ \citenamefont {Bloch}}]{lshaking1}%
  \BibitemOpen
  \bibfield  {author} {\bibinfo {author} {\bibfnamefont {Y.-A.}\ \bibnamefont
  {Chen}}, \bibinfo {author} {\bibfnamefont {S.}~\bibnamefont {Nascimb\`ene}},
  \bibinfo {author} {\bibfnamefont {M.}~\bibnamefont {Aidelsburger}}, \bibinfo
  {author} {\bibfnamefont {M.}~\bibnamefont {Atala}}, \bibinfo {author}
  {\bibfnamefont {S.}~\bibnamefont {Trotzky}}, \ and\ \bibinfo {author}
  {\bibfnamefont {I.}~\bibnamefont {Bloch}},\ }\href {\doibase
  10.1103/PhysRevLett.107.210405} {\bibfield  {journal} {\bibinfo  {journal}
  {Phys. Rev. Lett.}\ }\textbf {\bibinfo {volume} {107}},\ \bibinfo {pages}
  {210405} (\bibinfo {year} {2011})}\BibitemShut {NoStop}%
\bibitem [{\citenamefont {Krönke}\ \emph {et~al.}(2013)\citenamefont
  {Krönke}, \citenamefont {Cao}, \citenamefont {Vendrell},\ and\ \citenamefont
  {Schmelcher}}]{mlb1}%
  \BibitemOpen
  \bibfield  {author} {\bibinfo {author} {\bibfnamefont {S.}~\bibnamefont
  {Krönke}}, \bibinfo {author} {\bibfnamefont {L.}~\bibnamefont {Cao}},
  \bibinfo {author} {\bibfnamefont {O.}~\bibnamefont {Vendrell}}, \ and\
  \bibinfo {author} {\bibfnamefont {P.}~\bibnamefont {Schmelcher}},\
  }\href@noop {} {\bibfield  {journal} {\bibinfo  {journal} {New J. Phys.}\
  }\textbf {\bibinfo {volume} {15}},\ \bibinfo {pages} {063018} (\bibinfo
  {year} {2013})}\BibitemShut {NoStop}%
\bibitem [{\citenamefont {Cao}\ \emph {et~al.}(2013)\citenamefont {Cao},
  \citenamefont {Krönke}, \citenamefont {Vendrell},\ and\ \citenamefont
  {Schmelcher}}]{mlb2}%
  \BibitemOpen
  \bibfield  {author} {\bibinfo {author} {\bibfnamefont {L.}~\bibnamefont
  {Cao}}, \bibinfo {author} {\bibfnamefont {S.}~\bibnamefont {Krönke}},
  \bibinfo {author} {\bibfnamefont {O.}~\bibnamefont {Vendrell}}, \ and\
  \bibinfo {author} {\bibfnamefont {P.}~\bibnamefont {Schmelcher}},\
  }\href@noop {} {\bibfield  {journal} {\bibinfo  {journal} {J. Chem. Phys.}\
  }\textbf {\bibinfo {volume} {139}},\ \bibinfo {pages} {134103} (\bibinfo
  {year} {2013})}\BibitemShut {NoStop}%
\bibitem [{\citenamefont {Cao}\ \emph {et~al.}(2017)\citenamefont {Cao},
  \citenamefont {Bolsinger}, \citenamefont {Mistakidis}, \citenamefont
  {Koutentakis}, \citenamefont {Krönke}, \citenamefont {Schurer},\ and\
  \citenamefont {Schmelcher}}]{mlx}%
  \BibitemOpen
  \bibfield  {author} {\bibinfo {author} {\bibfnamefont {L.}~\bibnamefont
  {Cao}}, \bibinfo {author} {\bibfnamefont {V.}~\bibnamefont {Bolsinger}},
  \bibinfo {author} {\bibfnamefont {S.~I.}\ \bibnamefont {Mistakidis}},
  \bibinfo {author} {\bibfnamefont {G.~M.}\ \bibnamefont {Koutentakis}},
  \bibinfo {author} {\bibfnamefont {S.}~\bibnamefont {Krönke}}, \bibinfo
  {author} {\bibfnamefont {J.~M.}\ \bibnamefont {Schurer}}, \ and\ \bibinfo
  {author} {\bibfnamefont {P.}~\bibnamefont {Schmelcher}},\ }\href@noop {}
  {\bibfield  {journal} {\bibinfo  {journal} {J. Chem. Phys.}\ }\textbf
  {\bibinfo {volume} {147}},\ \bibinfo {pages} {044106} (\bibinfo {year}
  {2017})}\BibitemShut {NoStop}%
\bibitem [{\citenamefont {Meyer}\ \emph {et~al.}(1990)\citenamefont {Meyer},
  \citenamefont {Manthe},\ and\ \citenamefont {Cederbaum}}]{mctdh1}%
  \BibitemOpen
  \bibfield  {author} {\bibinfo {author} {\bibfnamefont {H.-D.}\ \bibnamefont
  {Meyer}}, \bibinfo {author} {\bibfnamefont {U.}~\bibnamefont {Manthe}}, \
  and\ \bibinfo {author} {\bibfnamefont {L.}~\bibnamefont {Cederbaum}},\
  }\href@noop {} {\bibfield  {journal} {\bibinfo  {journal} {Chem. Phys.
  Lett.}\ }\textbf {\bibinfo {volume} {165}},\ \bibinfo {pages} {73 } (\bibinfo
  {year} {1990})}\BibitemShut {NoStop}%
\bibitem [{\citenamefont {Beck}\ \emph {et~al.}(2000)\citenamefont {Beck},
  \citenamefont {J\"ackle}, \citenamefont {Worth},\ and\ \citenamefont
  {Meyer}}]{mctdh2}%
  \BibitemOpen
  \bibfield  {author} {\bibinfo {author} {\bibfnamefont {M.~H.}\ \bibnamefont
  {Beck}}, \bibinfo {author} {\bibfnamefont {A.}~\bibnamefont {J\"ackle}},
  \bibinfo {author} {\bibfnamefont {G.~A.}\ \bibnamefont {Worth}}, \ and\
  \bibinfo {author} {\bibfnamefont {H.~D.}\ \bibnamefont {Meyer}},\ }\href@noop
  {} {\bibfield  {journal} {\bibinfo  {journal} {Phys. Rep.}\ }\textbf
  {\bibinfo {volume} {324}},\ \bibinfo {pages} {1 } (\bibinfo {year}
  {2000})}\BibitemShut {NoStop}%
\bibitem [{\citenamefont {Alon}\ \emph {et~al.}(2008)\citenamefont {Alon},
  \citenamefont {Streltsov},\ and\ \citenamefont {Cederbaum}}]{mctdhb1}%
  \BibitemOpen
  \bibfield  {author} {\bibinfo {author} {\bibfnamefont {O.~E.}\ \bibnamefont
  {Alon}}, \bibinfo {author} {\bibfnamefont {A.~I.}\ \bibnamefont {Streltsov}},
  \ and\ \bibinfo {author} {\bibfnamefont {L.~S.}\ \bibnamefont {Cederbaum}},\
  }\href@noop {} {\bibfield  {journal} {\bibinfo  {journal} {Phys. Rev. A}\
  }\textbf {\bibinfo {volume} {77}},\ \bibinfo {pages} {033613} (\bibinfo
  {year} {2008})}\BibitemShut {NoStop}%
\bibitem [{\citenamefont {Alon}\ \emph {et~al.}(2014)\citenamefont {Alon},
  \citenamefont {Streltsov},\ and\ \citenamefont {Cederbaum}}]{mctdhb2}%
  \BibitemOpen
  \bibfield  {author} {\bibinfo {author} {\bibfnamefont {O.~E.}\ \bibnamefont
  {Alon}}, \bibinfo {author} {\bibfnamefont {A.~I.}\ \bibnamefont {Streltsov}},
  \ and\ \bibinfo {author} {\bibfnamefont {L.~S.}\ \bibnamefont {Cederbaum}},\
  }\href@noop {} {\bibfield  {journal} {\bibinfo  {journal} {J. Chem. Phys.}\
  }\textbf {\bibinfo {volume} {140}},\ \bibinfo {pages} {034108} (\bibinfo
  {year} {2014})}\BibitemShut {NoStop}%
\bibitem [{\citenamefont {Alon}\ \emph {et~al.}(2012)\citenamefont {Alon},
  \citenamefont {Streltsov}, \citenamefont {Sakmann}, \citenamefont {Lode},
  \citenamefont {Grond},\ and\ \citenamefont {Cederbaum}}]{mctdhx}%
  \BibitemOpen
  \bibfield  {author} {\bibinfo {author} {\bibfnamefont {O.~E.}\ \bibnamefont
  {Alon}}, \bibinfo {author} {\bibfnamefont {A.~I.}\ \bibnamefont {Streltsov}},
  \bibinfo {author} {\bibfnamefont {K.}~\bibnamefont {Sakmann}}, \bibinfo
  {author} {\bibfnamefont {A.~U.}\ \bibnamefont {Lode}}, \bibinfo {author}
  {\bibfnamefont {J.}~\bibnamefont {Grond}}, \ and\ \bibinfo {author}
  {\bibfnamefont {L.~S.}\ \bibnamefont {Cederbaum}},\ }\href@noop {} {\bibfield
   {journal} {\bibinfo  {journal} {Chem. Phys.}\ }\textbf {\bibinfo {volume}
  {401}},\ \bibinfo {pages} {2 } (\bibinfo {year} {2012})}\BibitemShut
  {NoStop}%
\bibitem [{\citenamefont {Jozsa}(1994)}]{fidelity}%
  \BibitemOpen
  \bibfield  {author} {\bibinfo {author} {\bibfnamefont {R.}~\bibnamefont
  {Jozsa}},\ }\href@noop {} {\bibfield  {journal} {\bibinfo  {journal} {J. Mod.
  Opt.}\ }\textbf {\bibinfo {volume} {41}},\ \bibinfo {pages} {2315} (\bibinfo
  {year} {1994})}\BibitemShut {NoStop}%
\bibitem [{\citenamefont {Will}\ \emph {et~al.}(2010)\citenamefont {Will},
  \citenamefont {Best}, \citenamefont {Schneider}, \citenamefont
  {Hackermuller}, \citenamefont {Luhmann},\ and\ \citenamefont
  {Bloch}}]{anticonfinement1}%
  \BibitemOpen
  \bibfield  {author} {\bibinfo {author} {\bibfnamefont {S.}~\bibnamefont
  {Will}}, \bibinfo {author} {\bibfnamefont {T.}~\bibnamefont {Best}}, \bibinfo
  {author} {\bibfnamefont {U.}~\bibnamefont {Schneider}}, \bibinfo {author}
  {\bibfnamefont {L.}~\bibnamefont {Hackermuller}}, \bibinfo {author}
  {\bibfnamefont {D.-S.}\ \bibnamefont {Luhmann}}, \ and\ \bibinfo {author}
  {\bibfnamefont {I.}~\bibnamefont {Bloch}},\ }\href@noop {} {\bibfield
  {journal} {\bibinfo  {journal} {Nature}\ }\textbf {\bibinfo {volume} {465}},\
  \bibinfo {pages} {197} (\bibinfo {year} {2010})}\BibitemShut {NoStop}%
\bibitem [{\citenamefont {Mazurenko}\ \emph {et~al.}(2017)\citenamefont
  {Mazurenko}, \citenamefont {Chiu}, \citenamefont {Ji}, \citenamefont
  {Parsons}, \citenamefont {Kanász-Nagy}, \citenamefont {Schmidt},
  \citenamefont {Grusdt}, \citenamefont {Demler}, \citenamefont {Greif},\ and\
  \citenamefont {Greiner}}]{anticonfinement2}%
  \BibitemOpen
  \bibfield  {author} {\bibinfo {author} {\bibfnamefont {A.}~\bibnamefont
  {Mazurenko}}, \bibinfo {author} {\bibfnamefont {C.~S.}\ \bibnamefont {Chiu}},
  \bibinfo {author} {\bibfnamefont {G.}~\bibnamefont {Ji}}, \bibinfo {author}
  {\bibfnamefont {M.~F.}\ \bibnamefont {Parsons}}, \bibinfo {author}
  {\bibfnamefont {M.}~\bibnamefont {Kanász-Nagy}}, \bibinfo {author}
  {\bibfnamefont {R.}~\bibnamefont {Schmidt}}, \bibinfo {author} {\bibfnamefont
  {F.}~\bibnamefont {Grusdt}}, \bibinfo {author} {\bibfnamefont
  {E.}~\bibnamefont {Demler}}, \bibinfo {author} {\bibfnamefont
  {D.}~\bibnamefont {Greif}}, \ and\ \bibinfo {author} {\bibfnamefont
  {M.}~\bibnamefont {Greiner}},\ }\href@noop {} {\bibfield  {journal} {\bibinfo
   {journal} {Nature}\ }\textbf {\bibinfo {volume} {545}},\ \bibinfo {pages}
  {462} (\bibinfo {year} {2017})}\BibitemShut {NoStop}%
\bibitem [{\citenamefont {Yang}\ \emph {et~al.}(2017)\citenamefont {Yang},
  \citenamefont {Dai}, \citenamefont {Sun}, \citenamefont {Reingruber},
  \citenamefont {Yuan},\ and\ \citenamefont {Pan}}]{spinDL}%
  \BibitemOpen
  \bibfield  {author} {\bibinfo {author} {\bibfnamefont {B.}~\bibnamefont
  {Yang}}, \bibinfo {author} {\bibfnamefont {H.-N.}\ \bibnamefont {Dai}},
  \bibinfo {author} {\bibfnamefont {H.}~\bibnamefont {Sun}}, \bibinfo {author}
  {\bibfnamefont {A.}~\bibnamefont {Reingruber}}, \bibinfo {author}
  {\bibfnamefont {Z.-S.}\ \bibnamefont {Yuan}}, \ and\ \bibinfo {author}
  {\bibfnamefont {J.-W.}\ \bibnamefont {Pan}},\ }\href@noop {} {\bibfield
  {journal} {\bibinfo  {journal} {Phys. Rev. A}\ }\textbf {\bibinfo {volume}
  {96}},\ \bibinfo {pages} {011602} (\bibinfo {year} {2017})}\BibitemShut
  {NoStop}%
\bibitem [{\citenamefont {Gagnon}\ \emph {et~al.}(2017)\citenamefont {Gagnon},
  \citenamefont {Fillion-Gourdeau}, \citenamefont {Dumont}, \citenamefont
  {Lefebvre},\ and\ \citenamefont {MacLean}}]{two-photon}%
  \BibitemOpen
  \bibfield  {author} {\bibinfo {author} {\bibfnamefont {D.}~\bibnamefont
  {Gagnon}}, \bibinfo {author} {\bibfnamefont {F.~m.~c.}\ \bibnamefont
  {Fillion-Gourdeau}}, \bibinfo {author} {\bibfnamefont {J.}~\bibnamefont
  {Dumont}}, \bibinfo {author} {\bibfnamefont {C.}~\bibnamefont {Lefebvre}}, \
  and\ \bibinfo {author} {\bibfnamefont {S.}~\bibnamefont {MacLean}},\
  }\href@noop {} {\bibfield  {journal} {\bibinfo  {journal} {Phys. Rev. Lett.}\
  }\textbf {\bibinfo {volume} {119}},\ \bibinfo {pages} {053203} (\bibinfo
  {year} {2017})}\BibitemShut {NoStop}%
\bibitem [{\citenamefont {Kastberg}\ \emph {et~al.}(1995)\citenamefont
  {Kastberg}, \citenamefont {Phillips}, \citenamefont {Rolston}, \citenamefont
  {Spreeuw},\ and\ \citenamefont {Jessen}}]{band-mapping1}%
  \BibitemOpen
  \bibfield  {author} {\bibinfo {author} {\bibfnamefont {A.}~\bibnamefont
  {Kastberg}}, \bibinfo {author} {\bibfnamefont {W.~D.}\ \bibnamefont
  {Phillips}}, \bibinfo {author} {\bibfnamefont {S.~L.}\ \bibnamefont
  {Rolston}}, \bibinfo {author} {\bibfnamefont {R.~J.~C.}\ \bibnamefont
  {Spreeuw}}, \ and\ \bibinfo {author} {\bibfnamefont {P.~S.}\ \bibnamefont
  {Jessen}},\ }\href@noop {} {\bibfield  {journal} {\bibinfo  {journal} {Phys.
  Rev. Lett.}\ }\textbf {\bibinfo {volume} {74}},\ \bibinfo {pages} {1542}
  (\bibinfo {year} {1995})}\BibitemShut {NoStop}%
\bibitem [{\citenamefont {F\"olling}\ \emph {et~al.}(2007)\citenamefont
  {F\"olling}, \citenamefont {Trotzky}, \citenamefont {Cheinet}, \citenamefont
  {Feld}, \citenamefont {Saers}, \citenamefont {Widera}, \citenamefont
  {Muller},\ and\ \citenamefont {Bloch}}]{band-mapping2}%
  \BibitemOpen
  \bibfield  {author} {\bibinfo {author} {\bibfnamefont {S.}~\bibnamefont
  {F\"olling}}, \bibinfo {author} {\bibfnamefont {S.}~\bibnamefont {Trotzky}},
  \bibinfo {author} {\bibfnamefont {P.}~\bibnamefont {Cheinet}}, \bibinfo
  {author} {\bibfnamefont {M.}~\bibnamefont {Feld}}, \bibinfo {author}
  {\bibfnamefont {R.}~\bibnamefont {Saers}}, \bibinfo {author} {\bibfnamefont
  {A.}~\bibnamefont {Widera}}, \bibinfo {author} {\bibfnamefont
  {T.}~\bibnamefont {Muller}}, \ and\ \bibinfo {author} {\bibfnamefont
  {I.}~\bibnamefont {Bloch}},\ }\href@noop {} {\bibfield  {journal} {\bibinfo
  {journal} {Nature}\ }\textbf {\bibinfo {volume} {448}},\ \bibinfo {pages}
  {1029} (\bibinfo {year} {2007})}\BibitemShut {NoStop}%
\bibitem [{\citenamefont {Cheinet}\ \emph {et~al.}(2008)\citenamefont
  {Cheinet}, \citenamefont {Trotzky}, \citenamefont {Feld}, \citenamefont
  {Schnorrberger}, \citenamefont {Moreno-Cardoner}, \citenamefont {F\"olling},\
  and\ \citenamefont {Bloch}}]{band-mapping3}%
  \BibitemOpen
  \bibfield  {author} {\bibinfo {author} {\bibfnamefont {P.}~\bibnamefont
  {Cheinet}}, \bibinfo {author} {\bibfnamefont {S.}~\bibnamefont {Trotzky}},
  \bibinfo {author} {\bibfnamefont {M.}~\bibnamefont {Feld}}, \bibinfo {author}
  {\bibfnamefont {U.}~\bibnamefont {Schnorrberger}}, \bibinfo {author}
  {\bibfnamefont {M.}~\bibnamefont {Moreno-Cardoner}}, \bibinfo {author}
  {\bibfnamefont {S.}~\bibnamefont {F\"olling}}, \ and\ \bibinfo {author}
  {\bibfnamefont {I.}~\bibnamefont {Bloch}},\ }\href@noop {} {\bibfield
  {journal} {\bibinfo  {journal} {Phys. Rev. Lett.}\ }\textbf {\bibinfo
  {volume} {101}},\ \bibinfo {pages} {090404} (\bibinfo {year}
  {2008})}\BibitemShut {NoStop}%
\end{thebibliography}%

\end{document}